\documentclass[prb, showpacs, preprint,preprintnumbers,amsmath,amssymb,eqsecnum,showkeys]{revtex4}

\usepackage{graphicx}
\usepackage{dcolumn}
\usepackage{bm}

\usepackage{xspace}
\newcommand{\ket}[1]{\ensuremath{|#1\rangle}\xspace}

\begin{document}

\preprint{}

\title{Comparison of strong coupling regimes in bulk GaAs, GaN and ZnO semiconductor microcavities}

\author{S. Faure}
 
 \email{faure@ges.univ-montp2.fr}

\author{T. Guillet}
\author{P. Lefebvre}
\author{T. Bretagnon}
\author{B. Gil}

\affiliation{Groupe d'Etude des Semiconducteurs (GES), Universit\'e Montpellier 2. CC 074. F-34095 Montpellier Cedex 5, France.}
\affiliation{CNRS, UMR 5650, F-34095 Montpellier, France}
\date{\today}

\begin{abstract}
Wide bandgap semiconductors are attractive candidates for polariton-based devices operating at room temperature. We present numerical simulations of reflectivity, transmission and absorption spectra of bulk GaAs, GaN and ZnO microcavities, in order to compare the particularities of the strong coupling regime in each system. Indeed the intrinsic properties of the excitons in these materials result in a different hierarchy of energies between the valence-band splitting, the effective Rydberg and the Rabi energy, defining the characteristics of the exciton-polariton states independently of the quality factor of the cavity. The knowledge of the composition of the polariton eigenstates is central to optimize such systems. We demonstrate that, in ZnO bulk microcavities, only the lower polaritons are good eigenstates and all other resonances are damped, whereas upper polaritons can be properly defined in GaAs and GaN microcavities.
\end{abstract}

\maketitle

\section{Introduction}

Semiconductor microcavities have recently attracted much attention because of the control that they provide on the light-matter interaction in solid state systems. In the strong coupling regime, excitons and photons form new coupled modes -the cavity polaritons- exhibiting large non-linearities which open the way to a broad area of fundamental and applied investigations. The most striking demonstrations have first been obtained on GaAs and CdTe-based microcavities: parametric oscillation \cite{Diederichs2006, Christopoulos2007}, parametric amplification \cite{Savvidis2000,Dasbach2000} and Bose-Einstein condensation \cite{Kasprzak2006}. However, those effects are obtained at cryogenic temperature, even though constant efforts are made and demonstrations of polariton emission in GaAs microcavities up to 200K have recently been reported \cite{Tsintzos2008}. The interest is now brought to wide band gap materials because the strong-coupling regime is stable up to room temperature \cite{Butte2006, Stokker-Cheregi2008} and the exciton binding energy is much larger, leading to stronger non-linearities. The realization of a GaN-based polariton laser operating at room temperature \cite{Christopoulos2007, Baumberg2008} has been achieved. A further enhancement in temperature is expected for ZnO-based microcavities, which development has just started \cite{Shimada2008, Vugt2006}.
Exciton binding energies (or effective Rydberg) and oscillator strengths in GaN and ZnO are about one order of magnitude larger than in GaAs. The resulting Rabi splitting in the strong coupling regime is equal to or larger than the effective Rydberg. Therefore, in GaN and ZnO, all exciton states are involved in the coupling to the photon mode. This different hierarchy of energies has a large impact on the spectroscopy of wide band gap microcavities, which cannot be understood by a simple shift of the standard description used for GaAs microcavities, i.e. a single exciton coupled to a single photon mode. In particular, the precise knowledge of the polariton eigenstates is crucial for a further modelling of the polariton scattering and the related non-linear effects.\\

This paper is intended to compare the polariton eigenstates and the optical spectra of comparable microcavities based on bulk GaAs, GaN and ZnO, i.e. $\lambda$-cavities with similar quality factors for the bare photon mode. We develop a standard approach based on transfer matrix and a quasi-particle model, including all the relevant terms that are of interest for wide bandgap materials. We compare the reflectivity spectra for the three investigated materials. We deduce in the case of ZnO the polaritonic eigenstates and show how to estimate their robustness. These simulations may be useful for comparison with experimental results obtained on GaAs, GaN or ZnO bulk microcavities.

We especially discuss what type of strong coupling can be expected in wide bandgap microcavities, and what quantity best describes the strength of this coupling. Indeed three ways exist to evidence the strong coupling regime in microcavities: (i) an energy splitting observable in reflectivity if oscillator strengths and photon spatial confinement are sufficiently strong; (ii) Rabi oscillations which are the temporal equivalent of the energy splitting and appear only if the lower and upper modes are well-defined, and finally (iii) phenomena like Bose-Einstein condensation or parametric scattering which require polaritons only on the lower polariton branch (LPB). This branch must also be well-defined and constitute a state with balanced exciton and photon contributions. The role of the microcavity is precisely to induce such a well-defined LPB. However we show below that the upper polariton branch (UPB) is not observable in ZnO-based microcavities since its coherence is damped by the absorption by scattering states of the exciton. In the last part of this paper, we investigate the role of the inhomogeneous broadening on optical spectra.

\section{Comparison between microcavities based on different materials}

Experimental reflectivity spectra of wide band gap semiconductor microcavities \cite{Butte2006, Shimada2008, Sellers2006, Antoine-Vincent2003} are dominated by inhomogeneous broadening and can appear more intricate than the simpler case of GaAs. In this work, we first investigate the intrinsic properties of bulk microcavities, with a small inhomogeneous broadening. \\
We compare different semiconductors in order to highlight their specificities and differences. To do this we choose three similar $\lambda$-cavities composed, for simplicity, of the same dielectric SiN/SiO$_{2}$ Bragg mirrors and where the active medium is constituted of bulk GaAs, GaN or ZnO, respectively, with a thickness that is tuned to induce appropriate resonances in the material, i.e. $d_{\lambda -cavity} = \lambda _{resonance} /N$, where $\lambda _{resonance}$ and $N$ are respectively the excitonic wavelength of the active medium and the refractive index of the material. These microcavities are virtual ones in the sense that it could be difficult or impossible in practice to realize their growth because substrates often differ from one material to another. But this approach is appropriate as we intend to compare intrinsic exciton-photon coupling for different semiconductors. Moreover the number of layers in the mirrors is adjusted to preserve the same quality factor \( (Q \sim 200) \) for the bare photon mode at the excitonic resonance. The spectrum of a bare cavity (a cavity without exciton resonances in the active layer) is calculated in order to determine the photon mode energy and its quality factor. Similar results are obtained for larger Q factors. Exciton resonances are attributed a homogeneous broadening of \( \gamma_{X}=0.3 \) meV in GaAs and \( \Gamma_{X}=1 \) meV in GaN and ZnO. The difference in the homogeneous broadening values results from the wish to preserve the same quality factor for excitons defined as \( Q=\frac{E_{g}}{\gamma_{X}}\simeq cte \)where $E_{g}$ is the band gap energy.\\

Reflectivity spectra are calculated for the three materials. The transfer matrix algorithm \cite{Born1999} is well adapted to the calculation of reflectivity and transmission spectra of microcavities. This algorithm requires calculating first the dielectric constant of the active medium.\\
In GaAs microcavities the active layer is described by one exciton in its fundamental state \cite{Tredicucci1995}. But the problem is more intricate in wurtzite GaN or ZnO due to the valence band structure and the strong oscillator strengths. In fact three parameters are important: the exciton binding energy or effective Rydberg (Ry*), the Rabi coupling energy $ \hbar \Omega_{Rabi}$ and the energy separation between excitons composed with different types of holes, i.e. the energy difference between the three valence band extrema. The hierarchy of these different energies changes significantly among the three considered materials and therefore the exciton-photon coupling will be affected in different ways. In bulk GaAs the two upper valence bands are degenerate \( (\Delta_{LH-HH} = 0 \ meV) \) and the Rabi splitting, for Q = 200, is smaller than the effective Rydberg (see Table \ref{tab:HoE}). As a result a simple anticrossing will be observed on angle-resolved reflectivity spectra. 
On the other hand, in wide band gap semiconductors the Rabi splitting energy, namely the exciton-photon coupling, can be of the same order of magnitude as (case of GaN) or much greater than (case of ZnO) the effective Rydberg and the energy separation between the three exciton bands. Here the anticrossing observed in angle-resolved reflectivity spectra will result from the coupling of the cavity mode with all the excitons simultaneously.\\

To take into account the hierarchy of energies of GaN and ZnO in the calculation of the dielectric constant we consider the fundamental (1s) A, B and C excitonic states as well as the ns excited states (whose oscillator strength is inversely proportional to \( n^{3} \), $n$ being the principal quantum number \cite{Elliott1957}). The scattering (continuum) states are also included. It is indeed necessary to take into account the absorption continuum in the simulations for two reasons: first, the Rabi splitting is larger than the effective Rydberg in wide band gap materials, pushing the UPB up to energies at which the continuum absorbs light. 
We must say that this situation can also be encountered in GaAs based microcavities, where the Rabi splitting and the effective Rydberg can be similar. In fact, what really makes a difference with GaN or ZnO is the product of the absorption coefficient by scattering states by the length of the cavity ($\lambda$-cavities in our situation). The data in Table \ref{tab:Comp} show that this product is at least ten times larger in GaN and ZnO than in GaAs and comparable to unity. To be more specific, a simple slab of ZnO having the same thickness of $\lambda /n$ would absorb, in a single pass, \(95\% \) of the photons having energies at the onset of the continuum. Therefore it is crucial to take into account the absorption by scattering states in the simulations for wide band gap semiconductors. \\

These physical differences being set, the excitons are described by classical oscillators, using an improved Drude-Lorentz model. The different states (fundamental and excited) are included as a sum of individual resonances in the absorption spectrum, each one with its own homogeneous and inhomogeneous broadening (in fact we take the same for all states). The absorption continuum due to scattering states is added and its amplitude adjusted to fit experimental absorption spectra. The parameters (oscillator strengths, exciton binding energy, etc.) for the different semiconductors were extracted from the literature (see \cite{Antoine-Vincent2003a, Mihailovic} and Table \ref{tab:HoE}, \ref{tab:Comp}  ).\\

The comparison between reflectivity spectra for GaAs, GaN and ZnO microcavities is shown in Fig.\ref{fig:Reflc}. The thickness of the active medium is adjusted to obtain the zero detuning condition (difference between the energy of the photon and of the lowest excitonic resonance) under normal incidence. In GaN and ZnO, only the A and B excitons have non-zero oscillator strength at normal incidence. Starting from the top panel, with the simple case of GaAs, we see two main dips, signature of the strong coupling regime. Dashed lines show the corresponding transmission spectrum.
By comparing the three material systems, we see that the splitting between polariton branches increases from GaAs to ZnO due to the increase of the oscillator strength. We also notice many structures that do not appear in the case of GaAs and are visible between the two extreme modes, making reflectivity spectra more intricate for wide band gap semiconductors. Finally the UPB is difficult to identify among all reflectivity dips as discussed in section III. Those ideal spectra strongly differ from the reported experimental results \cite{Tredicucci1995, Butte2006, Shimada2008} due to the main role played by the inhomogeneous broadening, as detailed in section V.\\
Figure \ref{fig:Refl1} presents a simulation of angle-dependent reflectivity spectra, for a fixed thickness of the active medium. All parameters used in calculation are the same as for Fig. 1, except that the zero detuning is here chosen to occur for an angle of $12^{\circ}$ (resp. $17^{\circ}$ and $25^{\circ}$) for the case of GaAs (resp. GaN and ZnO). GaAs presents polaritonic branches which broadening changes depending on the photon component since the broadening of the photon is larger than the one of the exciton. The case of GaN shows two successive anticrossings. The upper mode broadening at high angles is due to the crossing with the continuum of scattering states. In ZnO the angular dispersion of the LPB is well visible, whereas the upper mode becomes rapidly broader and fades with increasing angles due to the crossing with the continuum. Let us emphasise the fact that neither the broadening of the photon nor the one of the exciton are responsible for the broadening of the upper mode in ZnO microcavities. This broadening is solely related to the crossing of the mode with the continuum.

\section{Composition of the eigenstates}

We will now focus on the composition of the polaritonic states.
To determine the composition of the different states in terms of excitons and photons, the transfer matrix algorithm is no longer sufficient. We need to apply the quasi-particle model ~\cite{Savona1995}, which gives access to the homogeneous broadening of the eigenstates through the imaginary part of the energy and the expansion coefficients of the eigenstates on the exciton-photon basis. Due to the specificities of GaN and ZnO, we introduce in our quasi-particle matrix a quasi-continuum of states to account for the absorption by scattering states of the exciton. Let us note that our calculated spectra are displayed versus the incidence angle and not versus the in-plane wave-vector \( k_\parallel \) in order to be compared easily with experiments. However only the wave-vector is a good quantum number and is really used in the calculations as illustrated in Fig. \ref{fig:Refl1}: the yellow line shows the position of states with the same wave-vector. The ratio \( \hbar\Omega_{Rabi}/E_{g} \) being much greater in GaN and ZnO than in GaAs, the anti-crossing takes place over a wider angular domain (\( \sim 50^{\circ} \; \mathrm{instead \; of} \; 20^{\circ}) \) so that the angular dependence of the coupling now has to be included through the angular dependence of the oscillator strength. This effect is also very sensitive to the light polarization \cite{Kavokin2003}. Using the quasi-particle model, describing the eigenstates with \( k_\parallel \) instead of the angle \( \Theta \) alters the evaluation of exciton-photon coupling terms by a few percents.\\ 

The eigenstates given by the quasi-particle model in TE polarization (with the electric field perpendicular to the incidence plane) are shown in Fig. \ref{fig:Merge}(a-c) where the homogeneous broadening is represented in the form of error bars. For clarity, among all the possible excitonic states of the Balmer sequence, we have only included the 1s and 2s states. We can easily check that the quasi-particle description gives similar results as the transfer matrix method for the polariton branches. However, all the eigenstates (bright and dark) are visible in this representation. We can now give more details about the nature of each branch by comparing those results with reflectivity, transmission and absorption spectra (Fig.\ref{fig:RTA}). Indeed resonances observed in the absorption spectrum of the microcavity are related to incoherent interactions between the light and the excitonic states \cite{Agranovich2003}: at those energies, photons are simply absorbed by the active medium. On the contrary, resonances observed in the transmission spectra are highly sensitive to dephasing processes within the cavity \footnote{Transmission resonances can be seen as the constructive interference between multiple reflections in the Fabry-Perot cavity which have the same phase modulo $[2\Pi]$.} and correspond to the coherent interaction between photons and excitons, i.e. to the polaritons eigenstates deduced from the quasi-particle model. Reflectivity spectra are composed of both incoherent and polariton resonances, and should be interpreted carefully.\\

Let us first consider the case of GaAs: at resonance the LPB is composed for half of exciton $X_{1s}$ and half of photon (Fig.\ref{fig:Merge}-a,d). The two other states, labeled MPB (middle polariton branch) and UPB, are very close in energy. They both contribute to the UPB identified in experimental results \cite{Tredicucci1995}, for which the homogeneous broadening is three times larger than the one used in our model. All polariton branches appear in the transmission spectrum (Fig.\ref{fig:RTA}-d, g) and therefore constitute proper polariton states, even when the continuum is included in the calculation. However a strong incoherent absorption from excitons and the continuum is observed.
\\
In the case of GaN, we observe in the transmission spectrum (Fig.\ref{fig:RTA}-e,~h) one LPB corresponding to the eigenstate $\ket{1}$ and two strong MPBs (eigenstates $\ket{2}$ and $\ket{4}$) (Fig.\ref{fig:Merge}-b,~e). The UPB peak is much weaker than predicted from the photon content of eigenstate $\ket{5}$ \footnote{The validity of the quasi-particle model is reduced when the absorption is important, i.e., when $\alpha (E_{X}) \times d_{\lambda -cavity} \sim 1$; In transfer matrix the photon component of the polariton wavefunction is strongly different from the photon wavefunction in the bare cavity, whereas the exciton-photon basis of the quasi-particle model only includes the bare photon mode.}. The absorption above $3.5~eV$, mainly due to excitons, strongly damps the coherence of MPBs and UPB. Only the LPB remains a well defined polariton state. The impact of the continuum is not significant on the polariton branches.  
\\
In the case of ZnO (Fig.\ref{fig:Merge}-c,~f), the LPB at resonance is also composed for half of photon and half of excitons (essentially \( A_{1s} \) and \( B_{1s} \)). In the simple case where neither the excited states nor the continuum are considered and where the A and B excitons are degenerate in energy, this LPB state would be written:\\
\begin{center}
\( \ket{1}=\ket{LPB}=\frac{1}{2}(\ket{A_{1s}}+\ket{B_{1s}})+\frac{1}{\sqrt{2}}\ket{\nu} \),\\
\end{center}
which is very close to the composition calculated here. Using the same simple hypotheses as above the states labeled $\ket{2}$ and $\ket{4}$ (Fig. \ref{fig:Merge}-b) are dark (they are not coupled to the electromagnetic field) and would write:\\
\begin{center}
\( \ket{2}=\frac{1}{\sqrt{2}}(\ket{A_{1s}}-\ket{B_{1s}}) \).\\
\end{center}
When the continuum is not taken into account (Fig.\ref{fig:Merge}-i) the LPB and UPB appear as strong transmission peaks, whereas all MPBs are strongly damped by excitonic absorption. However, when the continuum is included in the calculation (Fig.\ref{fig:Merge}-f), the transmission of the UPB vanishes, and the asymmetric peak which is seen in reflectivity is simply the signature of the absorption by the continuum: the upper branch is no more defined. This confirms the results of the quasi-particle model: the UPB lies within the continuum spectral range and has a large broadening.\\
Let us now have a close look to the structures calculated in the intermediate spectral range, between 3.36 eV and 3.43 eV: the reflectivity dip at E = 3.415 eV is also strong in transmission, and corresponds to the middle polariton branch (MPB) that lies between the 1s and 2s exciton energies. All other structures are either weak or absent in the transmission spectra. The features at 3.377 eV and 3.42 eV (labeled $i_{1}$ and $i_{3}$) as well as those labeled $i_{2}$ and $i_{4}$ in Fig.\ref{fig:RTA}-i are related to the incoherent absorption by 1s and 2s excitons, and have been previously called "incoherent excitons" \cite{Agranovich2003}. They correspond to the excitonic resonances shown by the absorption coefficient (Fig.\ref{fig:RTA}-c) and indicated by dashed lines. The small dips at 3.37 eV and 3.386 eV arise from optical modes with a different order (labeled respectively as $3\lambda /2$ and $\lambda /2$ in Fig.\ref{fig:RTA}-i) of the cavity. They appear because the refractive index strongly varies in the vicinity of the excitonic gap, and therefore the same cavity can fulfil the resonance condition of a $\lambda $-cavity and a $3\lambda /2$ cavity for different energies.\\
The comparison between the three materials evidences their main differences: all polariton branches are well-defined eigenstates of the exciton-photon coupled system in GaAs bulk microcavities, whereas only lower polaritons subsist in GaN and ZnO microcavities. In order to compare their potentialities we have defined a figure of merit as the ratio, $ {\Delta E}/{\gamma_{LPB}} $, of the splitting in energy between the LPB and the next branch, over the homogeneous broadening of the LPB. This ratio is only worth 2 and 3 for GaAs and GaN, and increases to 10 for a ZnO microcavity. We should emphasize that we consider microcavities with the same quality factors for exciton and photon resonances. The calculated figure of merit are therefore larger than reported results for GaN and ZnO microcavities, but smaller than demonstrated in GaAs structures. For comparison, figure of merit of 20 to 50 can be obtained in GaAs and CdTe microcavities embedding quantum wells \cite{Balili2007,Deng2002, Bajoni2008,Kasprzak2006}. ZnO cavities are therefore attracting in order to increase this figure of merit.

\section{Estimation of the exciton-photon coupling}

The exciton-photon coupling term, V, is the parameter adjusted in the quasi-particle model and the one of interest for modelling the LPB as strongly coupled exciton and photon. For ZnO microcavities we obtain a coupling term V between the excitonic resonance and the cavity mode of V=100 meV, whereas the splitting between the LPB and the high energy mode observed in this simulation in the absence of continuum (\ref{fig:RTA}-i) is of 135 meV. This clearly differs from the case of GaAs (Fig.\ref{fig:RTA}-d) where the two dips in reflectivity arise from the LPB and the UPB, and the observed Rabi splitting represents directly the coupling term V. In the case of ZnO, the apparent splitting is in fact enhanced by the significant oscillator strength of excitonic excited states, which results in middle branches. The splitting between the LPB and the high energy mode is not therefore the relevant value in the sense that it is not representative of the exciton-photon coupling. 
Moreover due to the presence of the continuum states, the mode previously mentioned as the upper mode is in the weak coupling regime, even though the standard criterion for the strong coupling \cite{Savona1995} $V > \sqrt[]{\gamma_{X} \times \gamma_{\nu}}$, where $\gamma_{X}$ and $\gamma_{\nu}$ are the exciton and photon homogeneous broadening, is fulfilled. Hence no Rabi oscillations should be observable in bulk ZnO microcavities. Rabi oscillatons are not a sine qua non condition to observe the strong coupling regime, but only the result of two well-defined and well-balanced polariton states.

\section{Role of the inhomogeneous broadening}

The above mentioned sharp structures in the reflectivity spectra calculated nearby and above the exciton energy are the signature of the incoherent exciton absorption and the strong changes in the refractive index, which are both consequences of the large exciton oscillator strength. However they have never been seen experimentally in GaN microcavities, for example. Nevertheless our calculations remain consistent with the experimental results because we have left aside one important aspect: the inhomogeneous broadening. This broadening is important in current wide band gap microcavities for two reasons: (i) their growth is less well controlled than in more developed III-V and II-VI compounds, due to strain accumulation in the distributed Bragg mirrors, leading to structural defects; (ii) excitons in wide band gaps semiconductors have a small effective Bohr radius, so that the ratio of the Bohr radius over the length of the cavity \( d_{\lambda \! - \! cavity}/a_{B} \) (Table \ref{tab:Comp}) is at least twice smaller in wide band gap semiconductors. In GaAs based bulk microcavities, the absence of defects or strain fluctuations allows the fundamental exciton-polariton to have a well-defined energy, and possibly to enter in the so-called center-of-mass quantization regime \cite{Tredicucci1995} for states that are less rich in photon component. This is also favored by the averaging effect of a rather large Bohr radius. On the other hand, for wide band-gap materials, there exists more defects and potential fluctuations and the very small Bohr radii make it easier for the exciton-polariton to get trapped or at least scattered by potential fluctuations. The photon-like part of its wave-function still insures that the coherence length of the state is much larger than the thickness of the bulk layer, but this coherence length is significantly reduced. Experimentally the inhomogeneous broadening is greater in GaN and ZnO than in GaAs (a few ten meV instead of 1~meV in our simulations).\\
We compare the simulations of reflectivity spectra for the same ZnO $\lambda$ -cavity when varying the inhomogeneous broadening from \( \sigma = 1~meV \) to 10~meV and 30~meV (Fig. \ref{fig:ReflvsSigma}). The LPB is robust, but the intermediate dips attributed to higher order modes and incoherent excitons rapidly vanish. Finally, for $\sigma = 30 meV$, the MPB also disappears and the reflectivity spectrum simplifies into two well separated dips corresponding to the LPB and the continuum. Between these pics the reflectivity is worth $88\%$ due to the absorption by the swallowed resonances.

\section{Conclusion}

In summary we have presented numerical simulations of reflectivity, transmission and absorption spectra of bulk GaAs, GaN and ZnO microcavities to compare the specificities of the strong coupling regime in such systems. Wide band gap microcavities present a very specific hierarchy of interactions compared to GaAs or other II-VI microcavities: the expected Rabi splitting is larger than the effective Rydberg, which is in turn larger than the splitting between the A, B and C valence bands. Consequently excitonic excited bound and unbound states (absorption by scattering states above the band gap) must be taken into account. Several additional structures appear on reflectivity spectra as a result of the coupling of the cavity mode with these excitonic resonances. In the strong coupling regime this situation introduces middle polaritonic branches, not yet observed experimentally, due to the strong damping by excitonic absorption and the current inhomogeneous broadening in wide band gap microcavities. These middle branches induce a practical overestimation of the exciton-photon coupling term if one attemps to read it directly from reflectivity spectra. The so-called quasi-particle model enables to determine this coupling term. In the specific case of GaN and ZnO, we have demonstrated that only the lower polaritonic branch is a well-defined and well-mixed exciton-photon state, characterized by an intense transmission and a weak absorption. Finally, as a consequence of the broadening of the upper branch by the continuum, Rabi oscillations should not be observed in ZnO microcavities which nevertheless remain good candidates for polaritonic-based effects involving the lower polariton branch.\\
The authors acknowledge financial support of ANR under 'ZOOM' project ($n^{\circ} ANR-06-BLAN-0135$). 

\bibliographystyle{apsrev}

\bibliography{Biblio}

\newpage
\begin{widetext}

\begin{table}[htbp]
	\begin{tabular}{@{} l @{\quad} l @{\quad} l @{\quad} l @{}}
\hline
& GaAs & GaN & ZnO\\
 \hline
Rydberg (Ry*) & 4.8 meV & 27 meV &  60 meV \\
Valence band  splitting $(\Delta)$ & $\Delta_{LH-HH}=0 \ meV$ & $\Delta_{A1s-B1s}= 8 \ meV$ & $\Delta_{A1s-B1s}= 6 \ meV$ \\ 
 & & $\Delta_{A1s-C1s}= 26 \ meV$ & $\Delta_{A1s-C1s}= 50 \ meV$ \\
Rabi splitting $(\Omega_{Rabi})$ & 4 meV [\cite{Tredicucci1995}] & 30 meV [\cite{Antoine-Vincent2003}] & 120 meV  [\cite{Zamfirescu2002}]\\
 & $\Omega_{Rabi} < Ry^{*}$ & $\Omega_{Rabi}\cong Ry^{*} \cong \Delta$ & 
$\Omega_{Rabi} >> Ry^{*}, \Delta$ \\
\hline
	\end{tabular}\\
\caption{Hierarchy of energies and couplings in GaAs, GaN and ZnO.}
	\label{tab:HoE}

\end{table}

\begin{table}[htbp]
	\begin{tabular}{@{} l @{\quad} l @{\quad} l @{\quad} l @{}}
\hline
 & GaAs & GaN & ZnO\\
 \hline
Refractive index N & 3.46 & 2.6 & 1.9 \\
$\alpha (Eg) (cm^{-1})$ & $10^{4}$  [\cite{Sturge1962}] & $1.2 \times 10^{5}$  [\cite{Muth1997}] & $2 \times 10^{5}$  [\cite{Liang1968}] \\
$d_{\lambda -cavity}$ (nm) & 240 & 146 & 184 \\
$\alpha (Eg) \times d_{\lambda -cavity}$ & 0.2 & 1.8 & 3.7 \\
$a_{B}$ (nm) & 11 & 2.8 & 1.4 \\
$d_{\lambda -cavity}/ a_{B}$ & 22 & 52 & 131\\
\hline		
	\end{tabular}\\
\caption{Relevant parameters of GaAs, GaN and ZnO microcavities.}
	\label{tab:Comp}
\end{table}

\begin{center}
Figures
\end{center}

\begin{figure}[htbp]
   \begin{center}
      \includegraphics[width=0.5\linewidth]{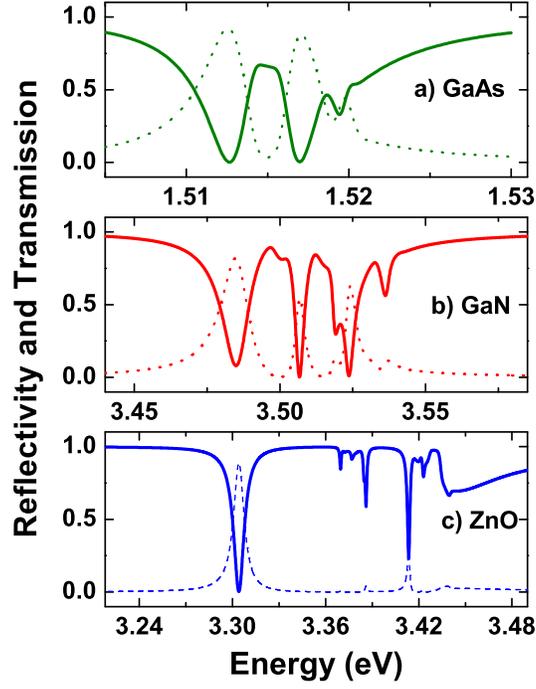}
   \end{center}
  
   \caption{Comparison of reflectivity (straight line) and transmission (dashed line) spectra simulations of GaAs, GaN and ZnO $\lambda$ -microcavities at zero detuning and normal incidence.}
  
   \label{fig:Reflc}
\end{figure}

\begin{figure}[hp]
   \begin{center}
      \includegraphics[width=1\linewidth]{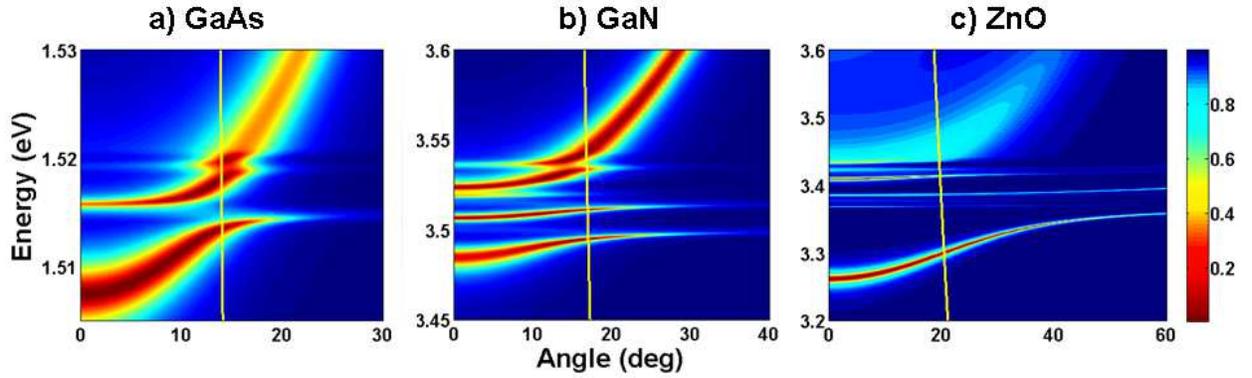}
   \end{center}
  
   \caption{(two columns wide, color online) Angle-resolved reflectivity spectra of $\lambda$-microcavities. The straight yellow line represents the states with the same wavevector.}
  
   \label{fig:Refl1}
\end{figure}

\begin{figure}[]
   \begin{center}
      \includegraphics[width=1\linewidth]{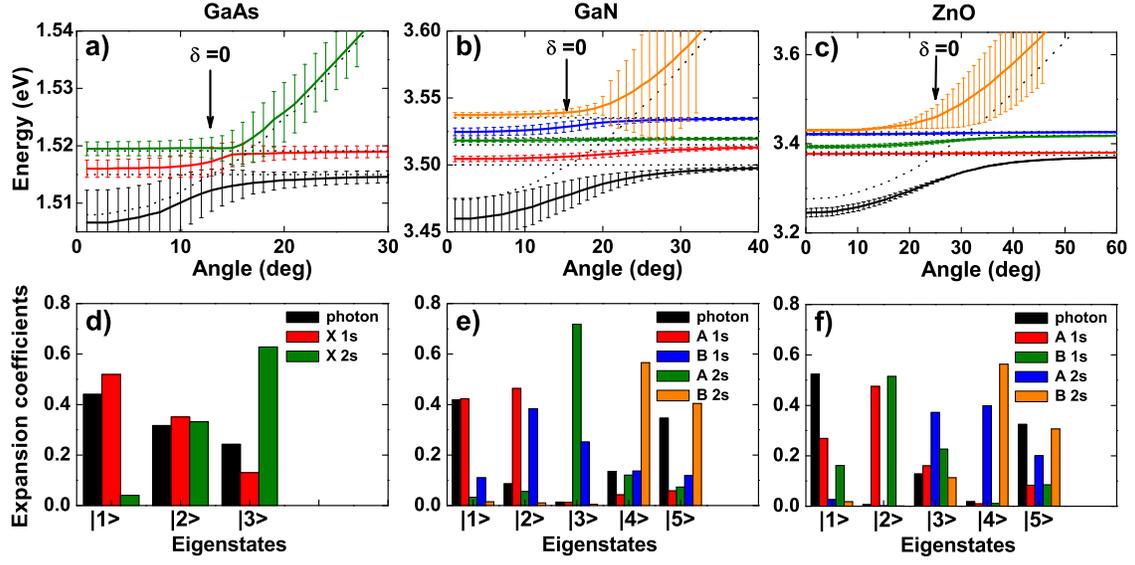}
   \end{center}
     
   \caption{(two columns wide, color online) a-c) Angle-resolved dispersion of eigenstates in $\lambda$-microcavities simulated with the quasi-particle model. This model provides the energies of the bright and dark eigenstates. The homogeneous broadening of each state is figured as error bar. Dashed lines indicate exciton and photon dispersions in the absence of coupling. d-f) Expansion coefficients of the eigenstates for the angle corresponding to the zero detuning within the exciton-photon basis.}
  
   \label{fig:Merge}
\end{figure}

\begin{figure}
		\includegraphics[width=1\linewidth]{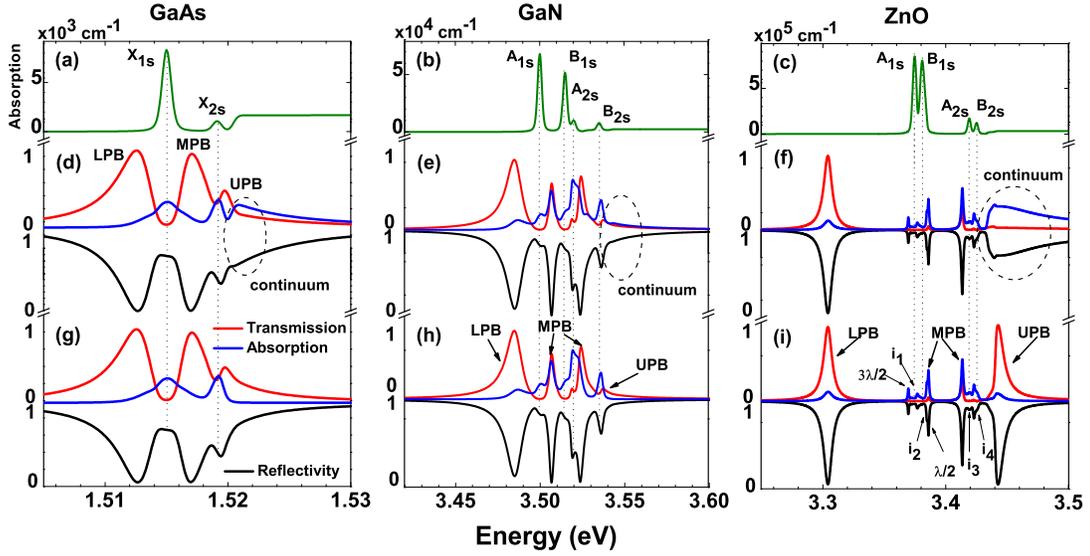}
\caption{(two columns wide, color online) a-c) ZnO Absorption coefficient showing the excitonic resonances, indicated by dashed lines on the others panels. Reflectivity, transmission and absorption spectra of a ZnO $\lambda -cavity$, calculated if the continuum is taken into account (d-f) or not (g-i).}
	\label{fig:RTA}
\end{figure}

\begin{figure}
	\begin{center}
		\includegraphics[]{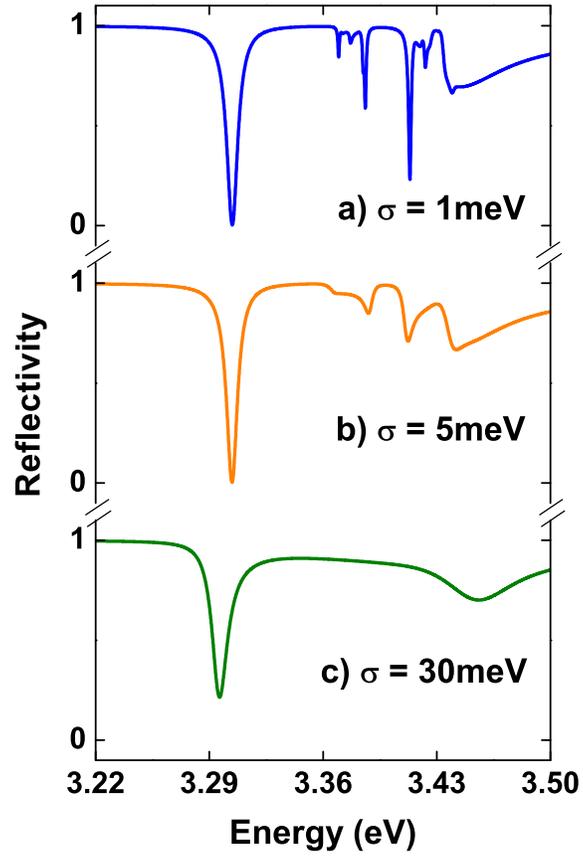}
	\end{center}
\caption{Reflectivity spectra of a ZnO $\lambda $-microcavity when varying the inhomogeneous broadening: a) $\sigma= 1 \ meV$, b) $\sigma= 5 \ meV$, c) $\sigma=30 \ meV$.}
	\label{fig:ReflvsSigma}
\end{figure}

\end{widetext}

\end{document}